\documentclass{aa} 

\usepackage{graphicx}

\usepackage{txfonts}
\usepackage{longtable}
\usepackage{amsmath}
\usepackage{graphicx}
\usepackage{float}
\usepackage{multirow}
\usepackage{subfigure}
\usepackage{rotfloat}
\usepackage{xcolor}
\usepackage{overpic}
\usepackage{caption}
\usepackage{soul}

\usepackage[]{hyperref}

\usepackage{breakurl}

\usepackage{array}
\usepackage[switch]{lineno} 
\usepackage{booktabs}
\usepackage{lscape}
\usepackage{subfigure}

\begin{document}

\title{GECAM discovery of a peculiar magnetar X-ray burst (MXB 221120) from SGR J1935+2154 associated with a fast radio burst}

\author{Wen-Jun~Tan\inst{1,2}
\and Yue~Wang\inst{1,2}
\and Chen-Wei~Wang\inst{1,2}
\and Shao-Lin~Xiong\inst{1}
\and Xiao-Bo~Li\inst{1}
\and Shuang-Nan~Zhang\inst{1,2}
\and Ce~Cai\inst{3}
\and Wang-Chen~Xue\inst{1,2}
\and Peng~Zhang\inst{1,4}
\and Bo-Bing~Wu\inst{1}
\and Zheng-Hua~An\inst{1}
\and Ming~Gao\inst{1}
\and Ming-Yu~Ge\inst{1}
\and Ke~Gong\inst{1}
\and Dong-Ya~Guo\inst{1}
\and Hao-Xuan~Guo\inst{1,5}
\and Long-Fei~Hao\inst{6,7}
\and Yue~Huang\inst{1}
\and Yu-Xiang~Huang\inst{6,2,7}
\and Ke-Jia~Lee\inst{8,9,6,10}
\and Bing~Li\inst{1}
\and Kui-Cheng~Li\inst{1}
\and Xin-Qiao~Li\inst{1}
\and Jia-Cong~Liu\inst{1,2}
\and Xiao-Jing~Liu\inst{1}
\and Ya-Qing~Liu\inst{1}
\and Xiang~Ma\inst{1}
\and Wen-Xi~Peng\inst{1}
\and Rui~Qiao\inst{1}
\and Yang-Zhao~Ren\inst{1,11}
\and Li-Ming~Song\inst{1,2}
\and Xi-Lei~Sun\inst{12}
\and Jin~Wang\inst{1}
\and Jin-Zhou~Wang\inst{1}
\and Ping~Wang\inst{1}
\and Xiang-Yang~Wen\inst{1}
\and Shuo~Xiao\inst{13,14}
\and Lun-Sheng~Xie\inst{15,1}
\and Heng~Xu\inst{9}
\and Sheng~Yang\inst{1}
\and Shu-Xu~Yi\inst{1}
\and Qi-bin~Yi\inst{16}
\and Zheng-Hang~Yu\inst{1,2}
\and Li-Da~Zhang\inst{1}
\and Fan~Zhang\inst{12}
\and Hong-Mei~Zhang\inst{12}
\and Jin-Peng~Zhang\inst{1,2}
\and Yan-Qiu~Zhang\inst{13,1,2}
\and Zhen~Zhang\inst{1}
\and Xiao-Yun~Zhao\inst{1}
\and Yi~Zhao\inst{17}
\and Chao~Zheng\inst{1,2}
\and Shi-Jie~Zheng\inst{1}
}

\institute{$^1$ State Key Laboratory of Particle Astrophysics, Institute of High Energy Physics, Chinese Academy of Sciences, Beijing 100049, China\\
\email{xiongsl@ihep.ac.cn}\\
$^2$ University of Chinese Academy of Sciences, Chinese Academy of Sciences, Beijing 100049, China\\
$^3$ College of Physics and Hebei Key Laboratory of Photophysics Research and Application, Hebei Normal University, Shijiazhuang, Hebei 050024, China\\
$^4$ College of Electronic and Information Engineering, Tongji University, Shanghai 201804, China\\
$^5$ Department of Nuclear Science and Technology, School of Energy and Power Engineering, Xi'an Jiaotong University, Xi'an 710049, China\\
$^6$ Yunnan Observatories, Chinese Academy of Sciences, Kunming 650216, China\\
$^7$ Key Laboratory for the Structure and Evolution of Celestial Objects, Chinese Academy of Sciences, Kunming 650216, China\\
$^8$ Department of Astronomy, Peking University, Beijing 100871, China\\
$^9$ National Astronomical Observatories, Chinese Academy of Sciences, Beijing 100101, China\\
$^{10}$ Beijing Laser Acceleration Innovation Center, Huairou, Beijing, 101400, China\\
$^{11}$ School of Physical Science and Technology, Southwest Jiaotong University, Chengdu 611756, China\\
$^{12}$ Institute of High Energy Physics, Chinese Academy of Sciences, Beijing 100049, China\\
$^{13}$School of Physics and Electronic Science, Guizhou Normal University, Guiyang 550001, China\\
$^{14}$Guizhou Provincial Key Laboratory of Radio Astronomy and Data Processing, Guizhou Normal University, Guiyang 550001, China\\
$^{15}$Institute of Astrophysics, Central China Normal University, Wuhan 430079, China\\
$^{16}$Key Laboratory of Lithium Battery New Energy Materials and Devices of Jiangxi Education Department, College of Intelligent Manufacturing and Materials \& Chemical Engineering, Yichun University, Yichun, Jiangxi Province, 336000, China.\\
$^{17}$School of Computer and Information, Dezhou University, Dezhou 253023, China}

\markboth{GECAM discovery of a peculiar magnetar X-ray burst (MXB 221120) from SGR J1935+2154 associated with a fast radio burst}{}

\abstract{Fast radio bursts (FRBs) are enigmatic cosmic transients of millisecond duration observed in the radio band.
The identification of FRB-associated magnetar X-ray bursts (MXBs) from galactic magnetar SGR J1935+2154 
suggests that at least a fraction of FRBs can be produced from magnetar activity. However, the sample size of FRB-associated MXBs is still very small.
Here we report a bright and peculiar FRB-associated MXB from SGR J1935+2154 detected by GECAM on November 20, 2022, dubbed MXB 221120.
We find that both temporal and spectral properties of MXB 221120 exhibit distinctive features. Its light curve could be generally described by a single FRED function with superposition of several narrow pulses. 
Interestingly, we identify a possible QPO feature with center frequency of $\sim$ 18 Hz in this MXB. The time-integrated spectrum is best fitted by a blackbody model with temperature ($kT$) of 18.6 keV, rendering it the first thermal spectrum FRB-associated MXB from SGR J1935+2154. Compared to other MXBs with single emission episode, MXB 221120 has longer duration and higher blackbody temperature, making it an outlier in the burst sample. These results indicate that MXB 221120 may be produced by a special mechanism with extreme physical conditions.}

\authorrunning{Tan et al.}
\titlerunning{FRB-associated MXB 221120}
\maketitle

\section{Introduction}

Magnetars are a unique type of neutron star (NS) which have typical spin periods of 2--12 s and spin-down rates of $\rm \sim 10^{-13}-10^{-10} s \,s^{-1}$. The energy emitted by magnetars is predominantly driven by magnetic fields \citep[e.g.][]{Ibrahim2004magnetar, Magnetar15Mereghetti} or magnetostatic energy \citep{Katz1982NS}. 
The (dipole) magnetic fields of magnetars range from $10^{14}$ to $10^{15}$ Gauss, making them hundreds to thousands of times stronger than those of ordinary NSs \citep[e.g.][]{Katz1982NS, Duncan92Magnetar, Thompson95SGR, Thompson96SGR}. 
Most of them are usually found in isolated system and sometimes associated with a supernova remnant \citep{Negro24magnetar}. 
The two most prominent manifestations of magnetars are soft gamma-ray repeaters (SGRs) and anomalous X-ray pulsars (AXPs) \citep{Mazets79asgr, Mazets79bsgr, Mereghetti95AXP, Magnetar08Mereghetti}. 
So far, about 30 magnetars have been discovered and confirmed, which are known in our Galaxy and in the Magellanic Clouds \citep{Cline82magnetar, Lamb02magnetar}. Although magnetars account for only a small fraction (less than 1$\%$) of the observed NS population \citep{Kaspi2017Magnetars}, they have played an important role in time-domain and multi-messenger astronomy.

The active galactic magnetar SGR J1935+2154 was first detected by Swift in 2014 \citep{swift14GCN} and lies within the supernova remnant G57.2+0.8, with a distance in the range of 4--13 kpc \citep{1935dis4, 1935dis12, 1935dis9}. SGR J1935+2154 has a spin period P = 3.24 s and a period derivative $ \dot{P} = 1.43\times  10^{-11} s \,s^{-1}$, implying a surface dipole magnetic field value of $ B \approx  2.2\times  10^{14}$ Gauss \citep{Israel2016sgr}.  

SGR J1935+2154 usually produces short magnetar X-ray bursts (MXBs), and observational properties of these bursts have been studied. The light curve pattern of MXBs from SGR J1935+2154 was found to be complicated, including two fundamental patterns called Fast Rise Exponential Decay (FRED) and Exponential Rise and Cut-Off Decay (ERCOD) \citep{ercod}. These bursts are likely to be generated by a kind of self-organizing critical process \citep{2023zhang1935}. The burst activity period was estimated to be $\sim127$ days \citep{2022xie1935}. Using one or two blackbody (BB) spectra to represent the thermal emission and a high-energy cutoff power-law (CPL) spectra for the non-thermal emission, the mean peak energy $E_{\rm p-CPL}$ of these bursts is about 26 keV, the mean soft BB temperature $ kT_{\rm BBs}$ is about 5 keV, and the mean hard BB temperature $ kT_{\rm BBh}$ is about 9 keV \citep{GBM_catalog_2021_Sep, GBM_catalog_2022_Jan, GBM_catalog_2022_Oct,HXMT1935cat, HXMT1935cat2, 2020LinSGR, GECAM_MXB_catalog}.

Magnetars are not only important objects in the X-ray band but also a crucial bridge connecting the X-ray and radio bands. Fast radio bursts (FRBs) are transient cosmic sources observed in the radio band that last only a few milliseconds \citep[e.g.][]{FRB_discovery, FRB_review}. Despite of many studies, their origin remains a mystery.
Proposed radiation mechanisms of FRBs could be generally classified into two categories: pulsar-like models \citep[e.g.][]{Zhang16frb, Lu20frb, yang18frb, Katz14frb, Lu18frb} and GRB-like models \citep[e.g.][]{Usov00frb, Lyubarsky14frb, Waxman17frb, Metzger19frb, Beloborodov20frb}. Both groups of models can be produced from magnetars \citep[e.g.][]{Popov10sgrfrb, Katz16sgrfrb, Wadiasingh20sgrfrb, Metzger17sgrfrb, Beloborodov17sgrfrb}.

On April 28, 2020, Canadian Hydrogen Intensity Mapping Experiment (CHIME) and Survey for Transient Astronomical Radio Emission 2 radio telescope (STARE2) detected a FRB, dubbed FRB 200428, generally localized at SGR J1935+2154 \citep{0428CHIME, START2_200428}. Meanwhile, several high-energy telescopes, including the Hard X-ray Modulation Telescope (\textit{Insight}-HXMT), detected an X-ray burst (dubbed MXB 200428 hereafter\footnote{It was tentatively named as XRB (acronym of X-ray burst) in literature, but XRB was already used to refer to X-ray binary. Therefore, we suggest to name it as magnetar X-ray burst (MXB) in accordance with the magnetar giant flare (MGF).}) from SGR J1935+2154 which is temporally and positionally associated with FRB 200428 \citep{HXMT0428gcn,HXMT0428, 0428integral, AGILE_0428, KW_0428}. The details of the light curve and expected dispersion delay indicate that the radio burst and the X-ray burst are the same burst event. Especially, \textit{Insight}-HXMT first reported that two narrow pulses in the X-ray burst are very likely the X-ray and hard X-ray counterpart of the FRB 200428 \citep{HXMT0428gcn}.
Therefore, FRB 200428 is the first galactic FRB, and the first FRB with high-energy electromagnetic counterpart. This discovery indicates that a fraction of the FRBs can be powered by magnetars. 
Remarkably, a $ 3.4\sigma$ quasi-periodic oscillation (QPO) feature at $\sim 40$~Hz has been identified in the light curve of MXB 200428 \citep{0428QPO}, which is a rare phenomenon among other MXBs from this magnetar.

In October 2022, SGR J1935+2154 became active again and produced a series of bursts. On October 14, 2022, both GECAM-B and GECAM-C were triggered in-flight \citep{GECAM_trigger,GECAM_alert} and from gound search ~\citep{GECAM_search_2025,ETJASMIN_II} by a bright MXB (designated MXB 221014) from SGR J1935+2154 which is associated with a FRB \citep{GECAM_221014_ATEL, CHIME_221014_ATEL, CHIME_221014_arxiv, GBT_221014_ATEL}. MXB 221014 represents the first GECAM detection of an FRB-associated MXB and the second case of MXB-FRB association just following MXB/FRB 200428~\citep{wang221014}. 

After GECAM discovery of MXB 221014, more MXBs (candidates) from SGR J1935+2154 were reported to be associated with radio bursts: A weak X-ray burst (MXB 221021) detected by \textit{Insight}-HXMT \citep{Atel1021HXMT} was reported to be associated with a radio burst discovered by Kunming 40-meter radio telescope (KM40m) at Yunnan Observatories \citep{Atel1021km40}.

In this paper, we report a bright and peculiar MXB from SGR J1935+2154 detected by GECAM-B on November 20, 2022, termed MXB 221120, which is associated with an FRB also from SGR J1935+2154 detected by the Kunming 40-meter radio telescope (KM40m) at Yunnan Observatories~\citep{Xiaobo221120}. This event is the second GECAM detection of an FRB-associated MXB. Here, we implemented detailed temporal and spectral analyses of this X-ray burst and made comparisons to other MXBs, especially those associated with FRB.

Throughout the paper, a standard cosmology with parameters $H_0$= 67.4 km \,$\rm{s^{-1}\,Mpc^{-1}}$, $\Omega_M$ = 0.315 and $\Omega_\Lambda$= 0.685 is adopted \citep{cosmology}. All parameter errors in this work are for the 68\% confidence level if not otherwise stated.

This paper is organized as follows. Observation and data reduction are described in Section \ref{section2}. Detailed temporal and spectral analyses are presented in Section \ref{section3}. A brief discussion is given in Section \ref{section4}, and summary is given in Section \ref{section5}.

\section{Observation and data reduction}\label{section2}

GECAM (Gravitational wave high-energy Electromagnetic Counterpart All-sky Monitor) constellation is composed of four instruments, GECAM-A/B (launched on December 10, 2020, \cite{xiao_GECAMB_time_calibration}), GECAM-C (launched on July 27, 2022, \cite{Zhang2023HEBS}) and GECAM-D (also known as
DRO/GTM, launched on March 14, 2024, \cite{Wang2024GTM, DROperformance_feng}), dedicated to monitoring all-sky gamma-ray transients. GECAM-A and GECAM-B feature a dome-shaped array of 25 Gamma-ray detectors (GRDs) and 8 Charged particle detectors (CPDs) while GECAM-C has 12 GRDs and 2 CPDs. All GRDs of GECAM-B operate in two readout channels: high gain (HG, 15--350 keV) and low gain (LG, 350--6000 keV). The dedicated design of GECAM facilitates detection in both time-domain astronomy and space environment \citep{OPP_wang} .

At 2022-11-20T11:25:04.850 (denoted as $T_0$, corresponding to MJD 59903.475751), GECAM-B was triggered in-flight \citep{GECAM_trigger,GECAM_alert} and from gound search \citep{GECAM_search_2025,ETJASMIN_II} by a bright MXB (denoted as MXB 221120) \citep{GECAM_MXB_catalog}. 
Both in-flight and gound search localization \citep{GECAM_localization,GECAM_MXB_catalog,ETJASMIN_I,ETJASMIN_II} are well consistent with SGR J1935+2154 within the error. At that time, GECAM-C, \textit{Insight}-HXMT and \textit{Fermi} were in the South Atlantic Anomaly (SAA), which highlighted the importance of GECAM-B observation.
Meanwhile, KM40m \citep{km40} detected a radio burst from SGR J1935+2154, which is temporally associated with MXB 221120~\citep{Xiaobo221120}. 

\begin{figure*}
\centering
\begin{tabular}{cc}
\multicolumn{2}{c}{\begin{overpic}[width=0.6\textwidth]{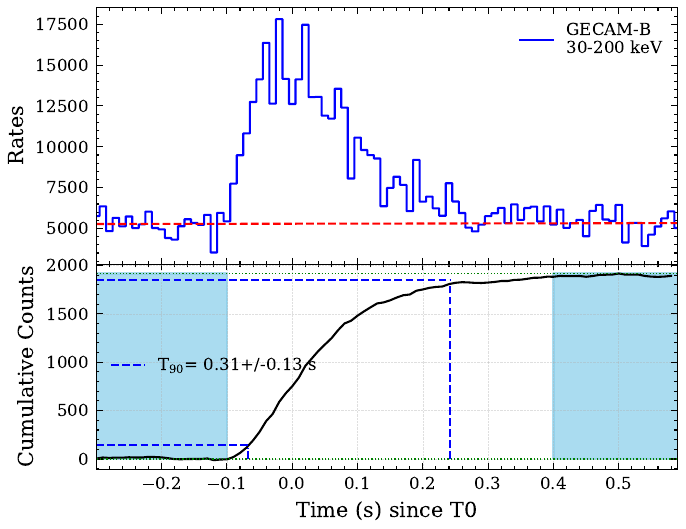}\put(4, 75){\bf a}\end{overpic}} \\
\begin{overpic}[width=0.35\textwidth]{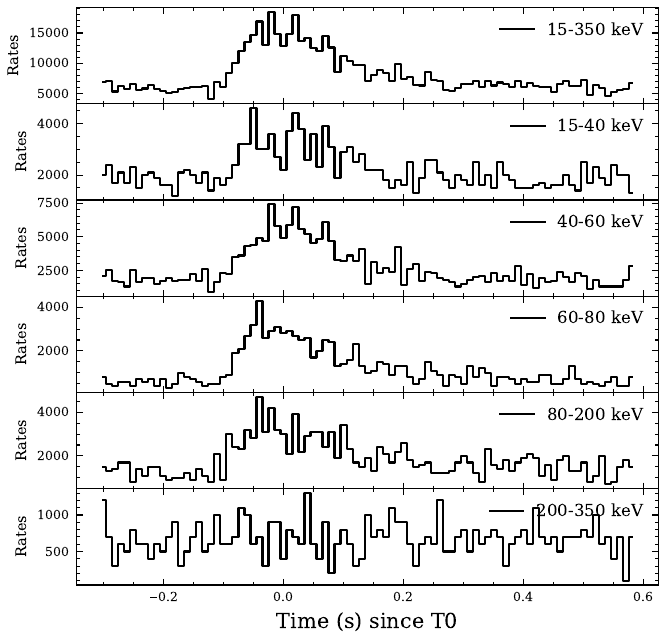}\put(0, 94){\bf b}\end{overpic} &
{\begin{overpic}[width=0.48\textwidth]{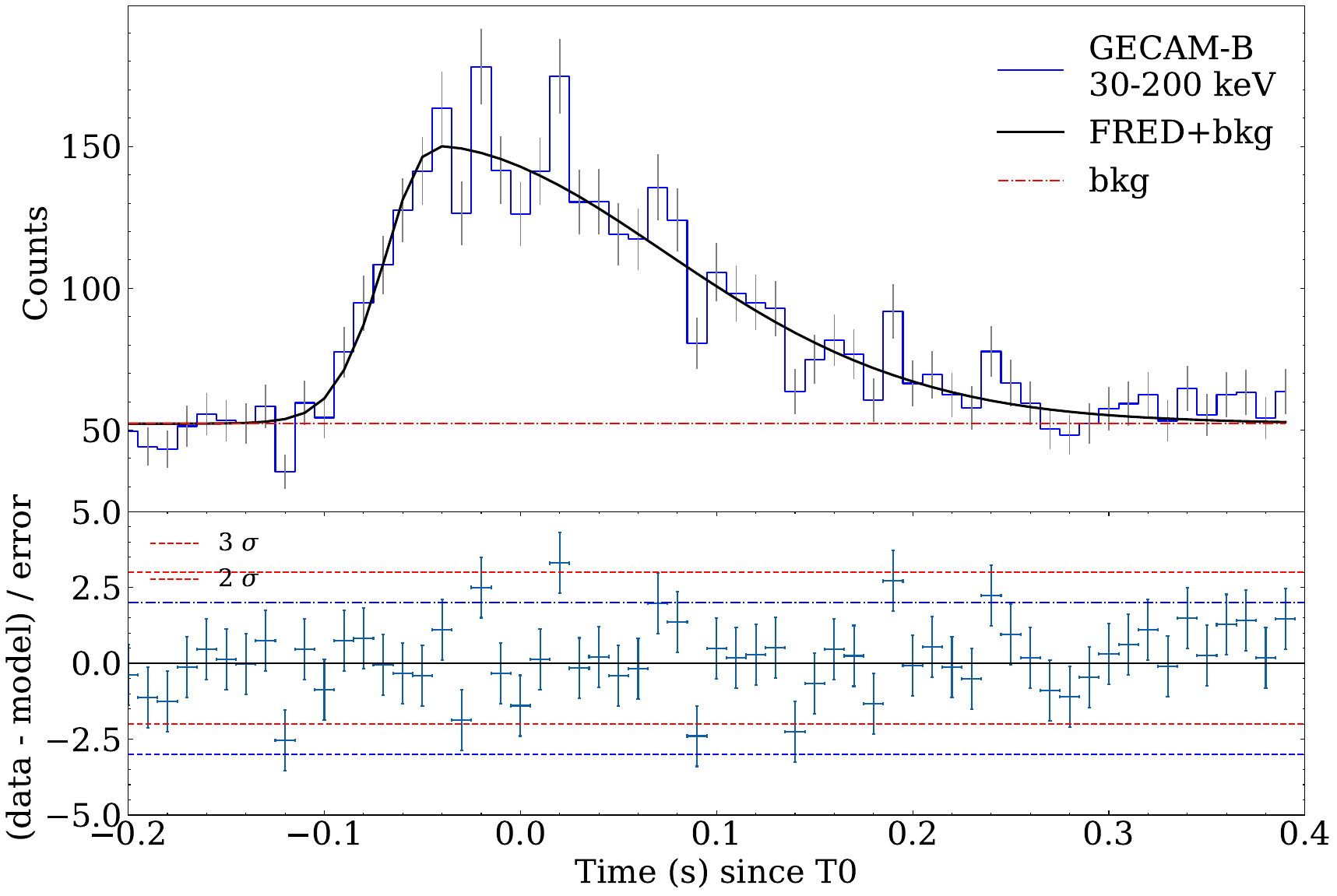}\put(0, 68){\bf c}\end{overpic}} \\
\end{tabular} 
\caption{\noindent\textbf{a}, Light curve in 10\,ms resolution and duration ($ T_{90}$). The red dashed line in top panel represents the background level. The blue shadow region in lower panel represents the background range. \textbf{b}, Light curves in different energy bands observed by GECAM-B.
\textbf{c}, Light curve fitting with FRED function. The light curve and FRED profile is plotted in top panel, and the residuals are plotted in lower panel.}
\label{fig:LC}
\end{figure*}

The light curve of MXB 221120 shows that this burst primarily consists of a single emission episode, i.e. a single FRED-shape pulse. The burst has a $ T_{90}$ (interval during which 90$\%$ of the X-ray burst fluence is detected) of 0.31 s (Fig.~\ref{fig:LC}a). 

Since significant signals were observed exclusively in the HG channel of GECAM-B (Fig.~\ref{fig:LC}b), only HG data are used for the data analysis.  
The temporal analysis was performed using detectors having an incident angle < 90$^\circ$ relative to the source (GRD 01, 02, 03, 04, 07, 08, 09, 10, 11, 17, 18, 19, 20, 25).
For spectral analysis, we utilized GECAMTools-v20240514 \footnote{\href{https://github.com/zhangpeng-sci/GECAMTools-Public}
{https://github.com/zhangpeng-sci/GECAMTools-Public}} to extract the spectra of the GRDs with incident angles smaller than 60$^\circ$, which are GRD 02, 03, 08, 09, 10, 18, 19 and 20. The energy range used for spectral fit is 30--350 keV considering the effective area and data quality.

\section{Data analyses and results}\label{section3}
\subsection{Temporal analysis}
\subsubsection{Light curve fitting}

Unlike MXB 200428, which consists of three broad bump-like components \citep{0428peak}, the light curve of MXB 221120 shows a single emission episode with several peaks on it. Therefore, it can be characterized by an overall profile to capture the general behavior. Here, we utilize a common FRED profile to fit the light curve \citep{1996fred}:

\begin{equation}
\label{eq:fred}
I(t) = 
\begin{cases} 
A \cdot e^{-\left(\frac{|t-t_{\rm max}|}{\sigma_1}\right)^\nu} & , t \leq t_{\rm max} \\
A \cdot e^{-\left(\frac{|t-t_{\rm max}|}{\sigma_2}\right)^\nu} & , t > t_{\rm max}
\end{cases}
\end{equation}
where $A$ is the normalization parameter, $t_{\rm max}$ is the peak time, $\sigma_{1}$ and $ \sigma_{2}$ are the rise and decay time, and $ \nu$ is the sharpness of the pulse. The fitting result is shown in Fig.~\ref{fig:LC}c.

By analyzing the light curves of a sample of MXBs \citep{GECAM_MXB_catalog} from SGR J1935+2154 detected by GECAM between 2021 and 2022, \cite{ercod} found that the light curve patterns of MXBs can be primarily classified into three categories: FRED, ERCOD and combination. These two fundamental patterns (FRED and ERCOD) of MXB exhibit distinct temporal and spectral characteristics \citep{ercod}. Therefore, we also tried to fit the light curve with ERCOD profile. The ERCOD function reads as: 
\begin{equation}
\label{eq:recd}
I(t) = 
\begin{cases} 
A \cdot e^{\left(-\frac{|t - t_{\rm max}|}{\sigma_1}\right)^\nu} & , t \leq t_{\rm max}, \\
A \cdot B \cdot\left[1 - e^{\left(-\frac{|t - t_{\text{cut}}|}{\sigma_2}\right)}\right] & , t_{\rm max} < t \leq t_{\text{cut}}
\end{cases}
\end{equation}
where $B = \left[1 - e^{\left(-\frac{|t_{\text{cut}} - t_{\rm max}|}{\sigma_2}\right)}\right]^{-1}$, and $ t_{\rm cut}$ is the cut-off time.

To test which profile is preferred, we use the Bayesian Information Criterion (BIC, \cite{1978BIC}) and reduced chi-square to perform the model selection. As shown in  Table~\ref{tab:fred}, the fitting of FRED profile to the light curve is significantly better than ERCOD.

\subsubsection{QPO search}

The significant QPO signal of $\sim$ 40 Hz found in MXB 200428 \citep{0428QPO} motivates us to check whether there is a similar QPO signal in MXB 221120.

\begin{figure*}
\centering
\begin{tabular}{cc}
\begin{overpic}[width=0.4\textwidth]{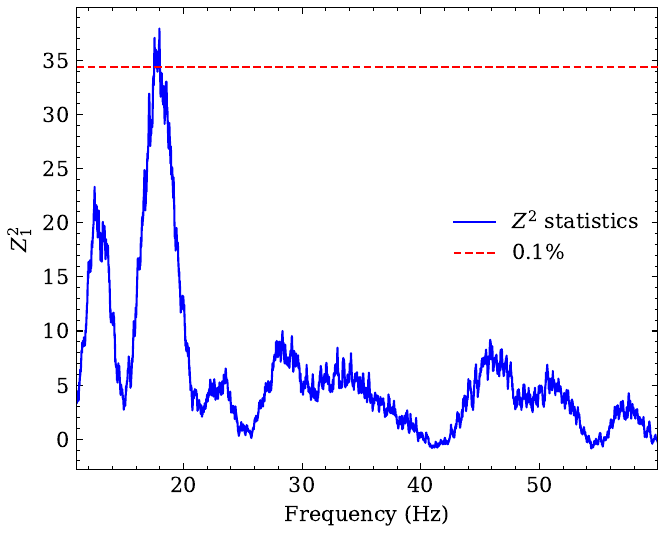}\put(2, 78){\bf a}\end{overpic} &
        \begin{overpic}[width=0.4\textwidth]{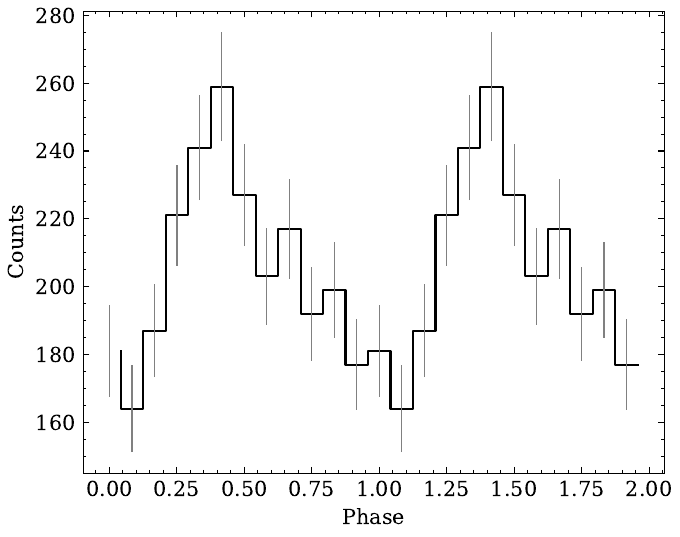}\put(2, 78){\bf b}\end{overpic} \\

\end{tabular}
\caption{\noindent\textbf{QPO searching of MXB 221120.} \textbf{a}, $Z_{1}^{2}$ periodogram of MXB 221120 
during the time interval of -0.1 s to 0.1 s. 
The red dashed line represents the false alarm probability of 0.1$\%$ level. \textbf{b}, Phase folding light curve of MXB 221120 with the best detected frequency.}
\label{fig:QPO}
\end{figure*}

We search for periodic signal through a $Z_{n}^{2}$-test procedure in a frequency range (10–60 Hz) around the known 40 Hz, with a step of 0.01 Hz and the number of harmonics n as 1, with the open-source Python software \textit{stingray} \citep{2019stingray,2024stingray}. During the time interval of $ T_{90}$, there is no significant signal in $ Z_{1}^{2}$ periodogram. However, when the time interval is limited to -0.1 s to 0.1 s, in which the signal is relatively high in the light curve, a significant signal that exceeds the false alarm probability of 0.1$\%$ in $ Z_{1}^{2}$ periodogram is located at F = $17.97_{-1.22}^{+1.56}$ Hz. The error is taken from the full width at half maxima (FWHM) of the peak of the $ Z_{1}^{2}$ curve. The $ Z_{1}^{2}$ periodogram and the false alarm probability level of 0.1$\%$ is shown in Fig.~\ref{fig:QPO}a. 
We further fold the burst photons in this time interval, using the best detected frequency (Fig.~\ref{fig:QPO}b). The phase light curve shows a sinusoidal structure with a significance of 4.2 $ \sigma$.

Interestingly, the obtained frequency is about half of
the common QPO (candidates) frequency $\sim 40$ Hz found in previous studies of MXBs \citep{0428QPO, 1935pds}. It is also worth noting that this frequency is somewhat close to the QPO frequency ($\sim$ 22Hz) found in the precursor of GRB 211211A, in which a magnetar is suggested to be involved in the compact binary merger \citep{11AQPO}.

Owing to the short duration and low frequency, the cycle number of the period is rather small ($\sim4$), therefore the significance of any potential low-frequency QPO would not be expected to be high, which is just the case that we find here.
As the feature of millisecond peaks distinguishes FRB-associated MXBs from normal MXBs \citep{Younes2020sgr, 0428peak}, and the peaks were also involved in the QPO of MXB 200428 \citep{0428QPO}, we consider that this possible QPO
of MXB 221120 maybe also contributed by the peaks superimposed on the broad component of the light curve (Fig.~\ref{fig:LC}a). 

\begin{table*}[htbp]
\caption{\centering{Fitting results of light curve}}
\begin{tabular*}{\hsize}{@{}@{\extracolsep{\fill}}c|ccccccc|cc}
\toprule
Model& &&& Parameters & & & &$ \chi^{2}_{\nu}$ (d.o.f) & BIC \\
& & & & & & & \\
 \hline
&A &$ t_{\rm max}$(ms)& $ \sigma_{1}$(ms) &$ \sigma_{2}$(ms)& $ \nu$ & bkg &&& \\
FRED& & & & & & & &\\
 & 49.5$^{+3.5}_{-3.1}$&-41.9$^{+5.2}_{-5.1}$ &33.7$^{+5.5}_{-5.1}$ &173.0$^{+11.4}_{-10.8}$ & 
 1.8$^{+0.3}_{-0.3}$&
 27.5$^{+1.1}_{-1.1}$ && 1.23 (114)& 168.76\\
 \hline
&A &$ t_{\rm max}$(ms)& $ \sigma_{1}$(ms) &$ \sigma_{2}$(ms)& $ \nu$ & bkg &$ t_{\rm cut}$(ms) \\
ERCOD& & & & & & & \\
 & 42.8$^{+2.2}_{-2.2}$&-13.0$^{+25.0}_{-22.0}$& 65$^{+26}_{-23}$&431$^{+51}_{-67}$& 4.1$^{+2.5}_{-1.5}$&28.3$^{+0.8}_{-0.8}$&165$^{+11}_{-12}$&1.42 (113)&194.22\\
\bottomrule
\end{tabular*}
\label{tab:fred} 
\end{table*}

\subsection{Spectral analysis}

We use Pyxspec software \citep{pyxspec} to perform spectral analysis. Past studies suggested that MXBs from SGR J1935+2154 can be described by phenomenological models such as a single blackbody (BB) or a double blackbody (BB plus BB) model for thermal emission, or a cutoff power-law (CPL) model for non-thermal emission \citep{2020LinSGR, GBM_catalog_2021_Sep,GBM_catalog_2022_Jan,GBM_catalog_2022_Oct}. For the time-integrated spectrum in the time interval of $ T_{90}$, 5 models are used, including BB model, BB plus power-law (PL) model, CPL model, CPL model with $\alpha$ fixed to -1 (corresponding to optically-thin thermal bremsstrahlung, namely OTTB), and double BB model. The PGSTAT
\footnote{\href{https://heasarc.gsfc.nasa.gov/docs/software/xspec/manual/node340.html}{https://heasarc.gsfc.nasa.gov/docs/software/xspec/manual/node340.html}} \citep{1996pgstat} is taken into account to test the goodness-of-fit.

The BB (bbodyrad in \textit{Xspec} \footnote{\href{https://heasarc.gsfc.nasa.gov/docs/software/xspec/manual/node137.html}{https://heasarc.gsfc.nasa.gov/docs/software/xspec/manual/node137.html}}) model reads as:
\begin{equation}   
N(E)=\frac{1.0344 \times 10^{-3}\times AE^2}{{\rm exp}(\frac{E}{kT})-1}
\label{equ:bb_Model}
\end{equation}
where $A=R^2_{\rm km}/D^2_{10}$ is the normalization constant, $R_{\rm km}$ is the source radius in km and $D_{10}$ is the distance to the source in units of 10 kpc, and $kT$ is the temperature in keV.

The PL model reads as:
\begin{equation}   
N(E) = A \left(\frac{E}{E_0}\right)^{\alpha},
\label{equ:pl_Model}
\end{equation}
where $A$ is the normalization constant ($\rm photons \cdot cm^{-2} \cdot s^{-1} \cdot keV^{-1}$), $\alpha$ is the power law photon index, and $E_0$ is the pivot energy fixed at 1 keV.

CPL model reads as:
\begin{equation}   
N(E)=A\left(\frac{E}{E_0}\right)^{\alpha} {\rm exp}(-\frac{E}{E_{\rm c}}),
\label{equ:CPL_Model}
\end{equation}
where $A$ is the normalization constant ($\rm photons \cdot cm^{-2} \cdot s^{-1} \cdot keV^{-1}$), $\alpha$ is the power law photon index, $E_0$ is the pivot energy fixed at 1 keV, and $E_{\rm c}$ is the characteristic cutoff energy in keV. The peak energy $E_{\rm p}$ is related to $E_{\rm c}$ through $E_{\rm p}$ = $(2 + \alpha)E_{\rm c}$.

\begin{figure}[http]
\includegraphics[width=0.44\textwidth]{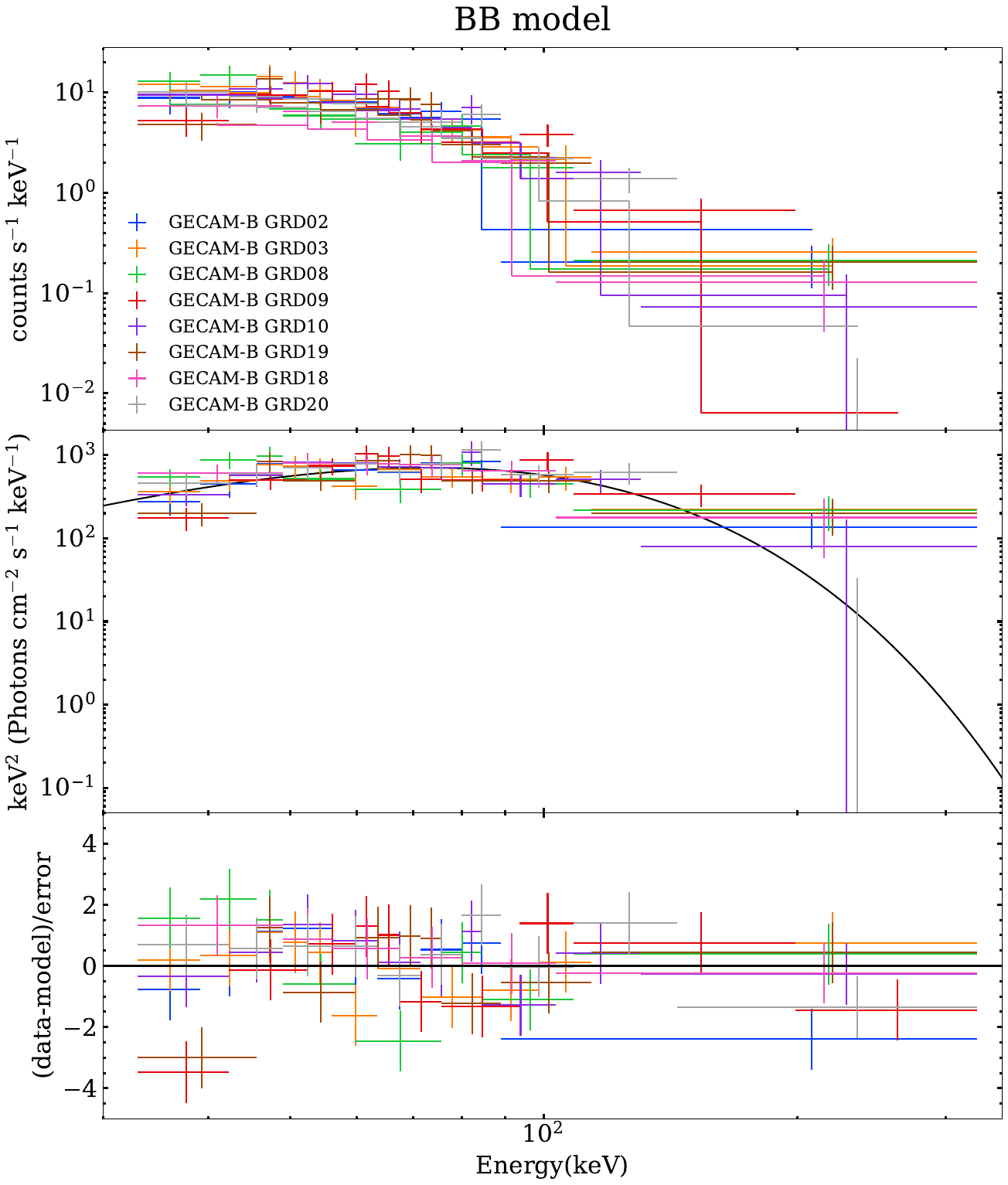}
\caption{The best spectrum fitting of MXB 221120.}
\label{fig:Spectrum}
\end{figure}

We use BIC \citep{1978BIC} to perform model comparison. For each model, BIC value is calculated and minimum BIC (BIC$\rm_{min}$) value is obtained. Models having $\Delta$BIC = BIC--BIC$\rm_{min}$ larger than 4 have considerably less support \citep{AIC_BIC}. Since the parameters of double BB model cannot be constrained, the results of the first four spectral models are listed in Table~\ref{tab:spectra}. Among these four models, BB has the minimum BIC, while $\Delta$BIC of BB plus PL model and CPL model with $\alpha$ fixed to -1 are greater than 4, indicating that these two models are not preferred. Furthermore, although the BIC of CPL is similar to that of BB, the $\alpha$ of CPL model is violently deviated from the typical value (-1), and is consistent with the index of a blackbody spectrum with Rayleigh-Jeans approximation in low-energy band, further proving that the best model is BB model (Fig~\ref{fig:Spectrum}).

Under the BB model, the burst fluence (in 35-300 keV) is $4.55^{+0.21}_{-0.14}\times 10^{-7}\, \rm erg\,\,\rm cm^{-2}$, with a total energy of $\sim$ 4.4$\times10^{39}$\,erg assuming the magnetar distance of 9 kpc \citep{1935dis9}. Moreover, based on the normalization of BB model, the emission region (radius) is calculated to be about 1.0 kilometer.

\begin{table*}[htbp]
\caption{\centering{The spectral results of MXB 221120}}
\begin{tabular*}{\hsize}{@{}@{\extracolsep{\fill}}cccccc}
\toprule
model & kT (keV)& $\alpha$ & $E_{peak}$(keV) & PGSTAT/d.o.f & BIC ($\Delta$BIC)\\
\hline
BB & 18.6$^{+0.5}_{-0.5}$ & - & -&292.44/206&303.11 (--)\\
BB+PL & 18.6$^{+0.6}_{-0.6}$ & -8.15$^{+2.9}_{-1.3}$ & - & 291.73/204&313.08 (9.97) \\
CPL & - & 1.97$^{+0.64}_{-0.5}$ & 72.8$^{+2.1}_{-2.0}$ & 288.62/205&304.63 (1.52) \\
CPL ($\alpha$ fixed to -1) & - & -1 & 65.6$^{+4.2}_{-3.9}$ & 350.24/206& 360.92 (57.81)\\
\bottomrule
\end{tabular*}
\label{tab:spectra}
\end{table*}

\subsection{MXB sample comparison}

\begin{figure*}
\centering
\begin{tabular}{ccc}
\begin{overpic}[width=0.31\textwidth]{ 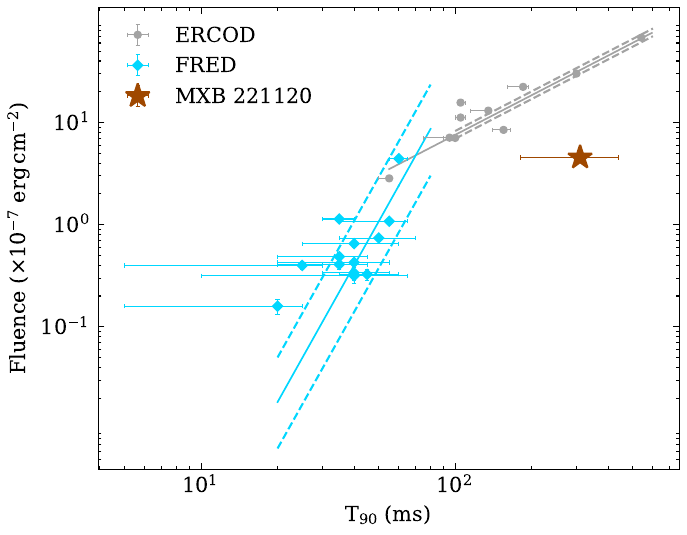}\put(5, 75){\bf a}\end{overpic} &
        \begin{overpic}[width=0.31\textwidth]{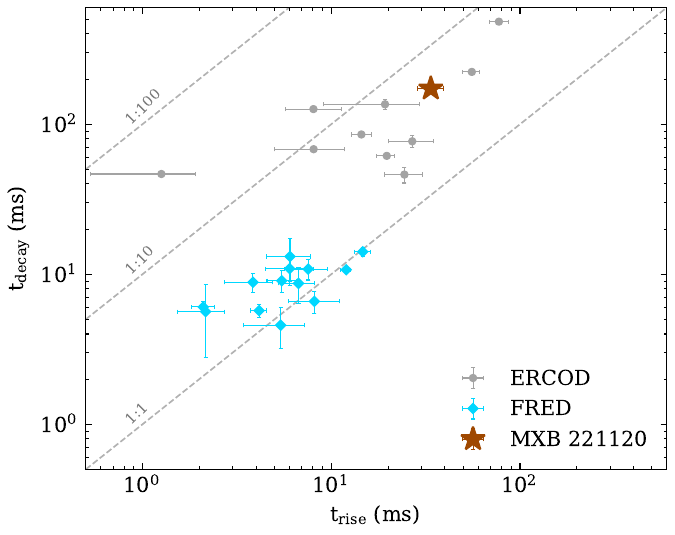}\put(2, 75){\bf b}\end{overpic} &
\begin{overpic}[width=0.31\textwidth]{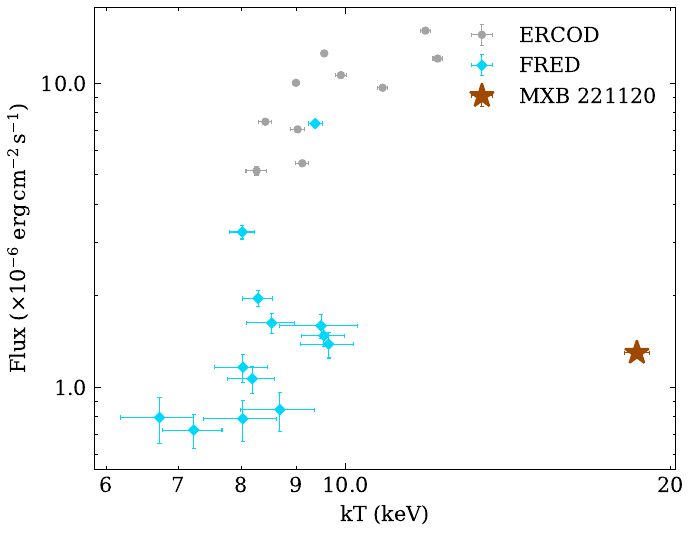}\put(5, 75){\bf c}\end{overpic} \\
        \begin{overpic}[width=0.31\textwidth]{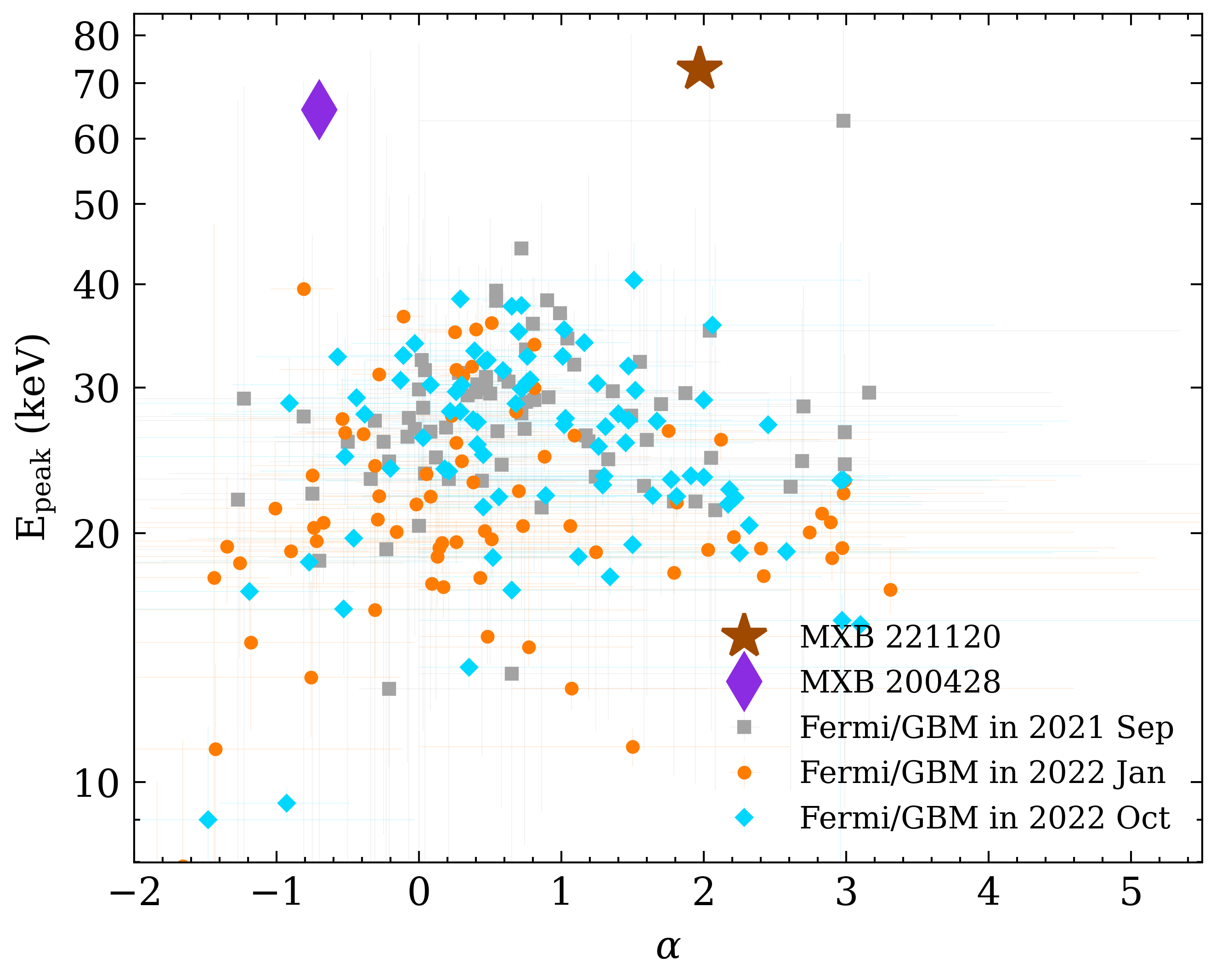}\put(2, 75){\bf d}\end{overpic} &
\begin{overpic}[width=0.31\textwidth]{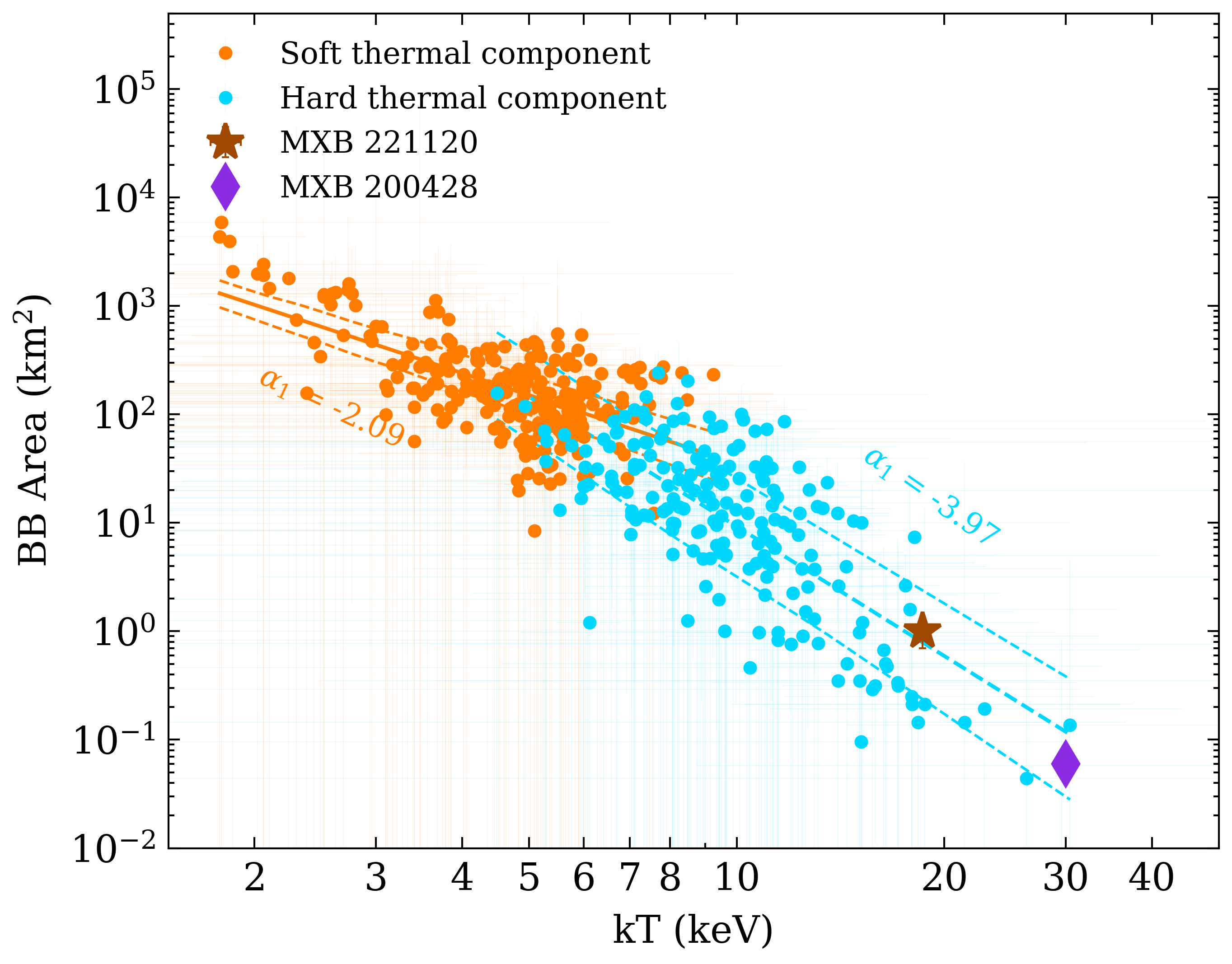}\put(5, 75){\bf e}\end{overpic}&
        \begin{overpic}[width=0.31\textwidth]{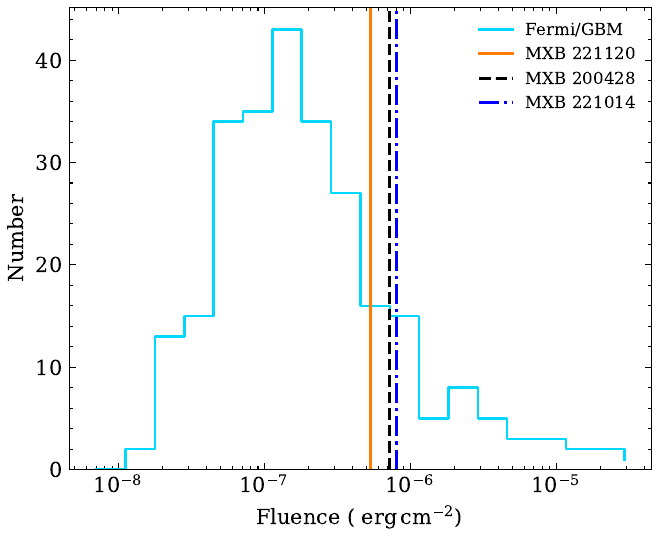}\put(2, 75){\bf f}\end{overpic} \\

\end{tabular}

\caption{\noindent\textbf{Several properties of MXB 221120 in sample distribution.} \textbf{a}, The position of duration ($ T_{90}$) and burst fluence in sample of FRED MXBs and ERCOD MXBs. The solid lines represent the power-law fitting of the data and the dashed lines represent the 1 $\sigma$ error region. \textbf{b}, The position of rise time and decay time in sample of FRED MXBs and ERCOD MXBs. The gray dotted lines indicate the ratio between rise time and decay time. \textbf{c}, The position of $kT$ and flux in sample of FRED MXBs and ERCOD MXBs. \textbf{d}, The position of $ \alpha$ and $ E_{\rm p}$ in \textit{Fermi}/GBM sample. \textbf{e}, The position of $kT$ and BB area in \textit{Fermi}/GBM sample. The solid lines represent the power-law fitting of the data and the dashed lines represent the 1 $\sigma$ error region. \textbf{f}, The position of fluence in \textit{Fermi}/GBM sample. In \textbf{a-c}, the sample is obtained from \cite{ercod}. In \textbf{d-f}, the sample is obtained from \cite{GBM_catalog_2021_Sep,GBM_catalog_2022_Jan,GBM_catalog_2022_Oct}}
\label{fig:Sample}
\end{figure*}

As presented above, MXB 221120 consists of a single emission episode, which can be perfectly described by a FRED profile. The light curve patterns of MXBs from SGR J1935+2154 are mainly categorized into FRED, ERCOD or a combination of these two fundamental patterns \citep{ercod}. MXBs with ERCOD shape generally exhibit longer durations and are brighter and harder than those with FRED shape \citep{ercod}. We systematically compare the properties of MXB 221120 with those of samples displaying FRED and ERCOD pattern. First, we fit the burst duration and burst fluence of both FRED and ERCOD patterns with a power-law function, respectively. Although MXB 221120 is well described by a FRED profile, its burst duration is an outlier of the FRED-pattern bursts (Fig.~\ref{fig:Sample}a), in particular, its rise and decay times are closer to those of the ERCOD-pattern bursts (Fig.~\ref{fig:Sample}b).

In addition to its temporal properties, the spectral features of MXB 221120 also represent an outlier. The $ kT_{\rm BB}$ of MXB 221120 is higher than those of all FRED-pattern bursts and ERCOD-pattern bursts (Fig.~\ref{fig:Sample}c), indicating a significantly harder fireball emission from MXB 221120. Furthermore, as the CPL model can also describe the burst, we compare the parameters of CPL model of MXB 221120 to other MXBs from SGR J1935+2154 detected by \textit{Fermi}/GBM from 2021 to 2022 \citep{GBM_catalog_2021_Sep,GBM_catalog_2022_Jan,GBM_catalog_2022_Oct}. We note that for the $E_{\rm p}$ of MXB 200428, we use the value of 65 keV from INTEGRAL/IBIS \citep{0428integral} rather than 37 keV from \textit{Insight}-HXMT \citep{HXMT0428}. The reason is that the result of \textit{Insight}-HXMT/HE is for a longer time period for the burst. If the time period around the peaks had been considered, a similar $E_{\rm p}$ to that of INTEGRAL/IBIS would have been obtained \citep{0428peak, HXMTperformance_zheng}. It can be seen that the $ E_{\rm p}$ of MXB 221120 is higher than all GBM samples as well as MXB 200428 (Fig.~\ref{fig:Sample}d).
 
The fitting of double BB for MXBs from SGR J1935+2154 facilitates the evaluation of the correlation between the BB emission areas (represented by the square of the source radius) and the $ kT_{\rm BB}$ \citep{2020LinSGR, GBM_catalog_2021_Sep,GBM_catalog_2022_Jan,GBM_catalog_2022_Oct}. We fit the BB model parameters (kT and emission area) of soft and hard components of all MXB samples detected by \textit{Fermi}/GBM during 2021 to 2022 using a PL function, yielding slopes of -2.09 for the soft components and -3.97 for the hard components, respectively (Fig.~\ref{fig:Sample}e). Interestingly, the $ kT_{\rm BB}$ of MXB 221120 is well consistent with the hard BB component region. This may indicate that MXB 221120 may still be in accordance with the double BB scenario. Due to the limitation of observation data or the much lower energy of the soft BB component, the soft BB component cannot be significantly detected. 
Notably, compared to all the samples from \textit{Fermi}/GBM, both the $ kT_{\rm BB}$ and the fluence of MXB 221120 are relatively large, but not outliers (see Fig.~\ref{fig:Sample}e and Fig.~\ref{fig:Sample}f). Indeed, it is worth noting that the fluence of all three FRB-associated MXBs (i.e. MXB 200428, MXB 221014, MXB 221120) is located in a rather narrow range of fluence (Fig.~\ref{fig:Sample}f). Whether this is intrinsic behavior or just coincidence is yet to be studied with more samples.
 
With these comparison studies, we find that while some properties of MXB 221120 fall within normal ranges among the burst sample (Fig.~\ref{fig:Sample}e and Fig.~\ref{fig:Sample}f), both the duration and $ kT_{\rm BB}$ of MXB 221120 are significant outliers among bursts characterized by a single process (either FRED-pattern or ERCOD-pattern, Fig.~\ref{fig:Sample}a, Fig.~\ref{fig:Sample}b and Fig.~\ref{fig:Sample}c).
These results make the MXB 221120 a very peculiar burst and provide important implications on the physical processes.

\section{Discussion}\label{section4}

MXB 221120 was discovered at the tail end of the third active period of SGR J1935+2154 after MXB 200428. In contrast to the first two active periods (September 2021 and January 2022) during which no FRB-associated MXB was reported, in the third active period (October 2022) several FRB-associated MXBs were reported, including those produced after the ending of this active period~\citep{GECAM_221014_ATEL, CHIME_221014_ATEL, CHIME_221014_arxiv, GBT_221014_ATEL, Atel1021HXMT, Atel1021km40, GBM_catalog_2022_Oct}.

Compared to the bursts observed in October 2022, MXB 221120 exceeds the mean burst duration by a factor of $\sim3$, while the fluence is comparable to the mean value \citep{GBM_catalog_2022_Oct}. The $E_{\rm p}$ of MXB 221120 is about three times higher than the mean $E_{\rm p}$ of these bursts, while the $ kT_{\rm BB}$ is also higher than the mean value \citep{GBM_catalog_2022_Oct}. The longer duration and harder spectrum of MXB 221120 are consistent with that of MXB 200428, compared to other bursts in their respective active period \citep{GBM_catalog_2022_Oct, Younes2021magnetar}. These characteristics may represent the common properties that distinguish FRB-associated MXBs from ordinary MXBs, with previous studies also noting that waiting times during FRB episode are significantly shorter compared to other active periods \citep{Xie2024ApJ}.

From 2020 to 2022, both $E_{\rm p}$ and hard $ kT_{\rm BB}$ of bursts from SGR J1935+2154 exhibited a slight softening trend \citep{GBM_catalog_2022_Oct}. 
However, the $E_{\rm p}$ of MXB 221120 is considerably higher, even higher than that of MXB 200428 (Fig.~\ref{fig:Sample}d). Rather than attributing this discrepancy to physical origins, we propose that it is more likely due to data limitations. The deficiency of lower energy band data of GECAM-B may lead to an overestimation of $E_{\rm p}$ in MXB 221120. This conclusion is supported by the fact that the maximum effective plasma temperature ($kT_{\rm e,max}$) of about 24 keV, derived from $E_{\rm p} \sim 3\,kT_{\rm e,max}$ \citep{Thompson95SGR, 2020LinSGR}, is significantly higher than $ kT_{\rm BB}$.

A possible QPO signal is observed in this burst, with the center frequency being 17.97 Hz, about half of the 40 Hz QPO in MXB 200428 \citep{0428QPO}. The most common explanation of QPO signal is the torsional oscillations of the magnetar crust \citep{Negro24magnetar}, and the possible QPO signal of $\sim$ 18 Hz in MXB 221120 is consistent with a low-order crustal torsional eigenmode of a magnetar \citep{Duncan1998qpo, Wadiasingh2020QPO}. On the other hand, in the case of MXB 200428, millisecond peaks aligning with the FRB pulses are involved in the oscillation, therefore this possible QPO signal in MXB 221120 may also be related to narrow peaks. \cite{wu2025frb} proposed that these narrow X-ray peaks can be attributed to the inverse Compton scattering (ICS) of FRB photons by an extreme pair flow around the light cylinder, in which the pair flow possible originates from the compression and acceleration due to rapid rearrangement of the magnetic field and magnetic reconnection in the current sheet. In this scenario, the QPO in X-ray peaks may be caused either by the FRB photons or by extreme pair flow, and the essence of the oscillation is attributed to the regular evolution of the magnetosphere.

We note that MXB 221120 is the first thermal spectrum MXB associated with a FRB from SGR J1935+2154, which is best fitted by a BB model with the temperature of 18.6 keV, exhibiting a significant proportion of thermal radiation. Notably, the thermalization is universal for normal MXB, and the temperature of about 20 keV is consitent with the theoretical prediction \citep{Katz1996sgr}. However, for FRB-associated MXBs, MXB 221120 is the first case that is dominated by thermal emission.
We stress that this thermal spectrum contradicts previous observations \citep{HXMT0428, 0428integral}, and challenges current theoretical models \citep{Younes2021magnetar, Yang2021FRB, lin2011mxb}, which suggested that the FRB-associated MXBs are dominated by non-thermal components.
To check whether there is purely thermal emission in this burst, we estimate the photon number density of this burst by $n_{\gamma}\sim {\cal F}d^2/(R^2c)$, where ${\cal F}$ is the burst flux per $kT$, $d$ is the source distance of 9 kpc, and $R$ is the emission region size \citep{2020LinSGR}. The $ n_{\gamma}$ of MXB 221120 is $\rm \sim 1.1\times 10^{26} cm^{-3}$, which is comparable to the photon density of a purely Planckian distribution $0.24\, (kT/[\lambda_{c}\,m_ec^2])^3 \sim 2\times10^{26}$cm$^{-3}$ for a reduced electron Compton wavelength $\lambda_{c} = \hbar/m_ec$ \citep{2020LinSGR}. This result indicates a perfect thermalization of this burst emission.

Normal MXBs are usually suggested to be produced from the rapid release of magnetic energy when the sudden cracking of the magnetar crust is induced by the twisted magnetic field. The released energy excites plasma, generating a photon pair plasma bubble confined by closed magnetic field lines, referred to as a ``trapped fireball". This trapped fireball 
cools and shrinks through X-ray emissions observed as MXB bursts. Due to the high opacity of plasma and local thermodynamic equilibrium, the emissions are thermalized, as reflected in the burst spectra \citep[e.g.][]{Thompson95SGR, Cheng1996magnetar, Younes2014sgr, Katz1996sgr}. 

For FRB-associated MXBs, it is believed that after the FRB is emitted, the electron plasma in the magnetar magnetosphere would be accelerated or decelerated by the radiation force to reach an equilibrium state. The X-ray spectrum is then altered by resonant Compton scattering of these balanced relativistic electrons, resulting in a high-energy cutoff, just like MXB 200428 \citep{Yang2021FRB, Yamasaki2020magnetar}. 

In MXB 221120, the dominant BB spectrum suggests that thermal photons escape from the trapped fireball with a size of about 1 km, without undergoing resonant Compton scattering. Two possible scenarios could explain why scattering may not occur: one is that the electron density in the magnetosphere is relatively low, resulting in insufficient scattering; the other is that the radiation emitted by the fireball is not spatially co-located with the high-density electrons in the magnetosphere. 

As an event characterized by a single process (namely a single FRED shape), MXB 221120 exhibits unusually long duration and high kT compared to other MXBs characterized by the same single process. From the perspective of its association with FRB, this is a natural result of a burst occurring near the polar cap region. Since FRBs are more likely to be triggered near polar caps \citep{Beloborodov2009magnetar, Yang2021FRB, GBM_catalog_2022_Oct}, a dichotomy arises between FRB-associated MXBs and ``orphan" MXBs, with the former occurring in quasi-polar regions and the latter occurring far from the polar caps \citep{Younes2021magnetar}. The strong local magnetic fields near the polar cap lead to tighter confinement of the fireball, resulting in a higher temperature and smaller emission area, consistent with observations (Fig.~\ref{fig:Sample}e). Thus, as an FRB-associated MXB, MXB 221120 is distinctive in that its emission exhibits a single emission process lasting for an extended duration. Given that the trigger mechanism for MXBs is analogous to earthquakes \citep{Thompson95SGR, Cheng1996magnetar}, we suggest that this burst is likely caused by a single crust crack.

\section{Summary}\label{section5}

MXB 221120 is a bright and peculiar FRB-associated MXB detected by GECAM-B from SGR J1935+2154. In this study, we comprehensively analyzed this burst, found interesting temporal and spectral properties, and discussed physical implications and interpretations.

In this burst, we find a possible QPO signal with the center frequency of about 18 Hz, which is about half of the 40 Hz QPO found in MXB 200428 \citep{0428QPO}. We suggest that this possible QPO can be attributed to a low-order crustal torsional eigenmode of a magnetar \citep{Duncan1998qpo, Wadiasingh2020QPO}, similar to that in MXB 200428 \citep{0428QPO}. Additionally, this QPO may be phenomenologically related to narrow pulses, suggesting that it may reflect the regular evolution of the magnetosphere \citep{wu2025frb}.

We also find that this burst is best fitted by a BB model with temperature of 18.6 keV, rendering it the first thermal spectrum FRB-associated MXB from SGR J1935+2154. This finding contradicts previous observations \citep{HXMT0428, 0428integral}, and 
challenges current theoretical models \citep{Younes2021magnetar, Yang2021FRB}, which propose that the FRB-associated MXBs are dominated by non-thermal components resulting from resonant Compton scattering. However, the thermal spectrum indicates that photons escape from the trapped fireball without undergoing scattering either due to the relatively low electron density in the magnetosphere, resulting in insufficient scattering, or because the radiation from the fireball is not spatially co-located with the high-density electrons in the magnetosphere. Regardless of the scenario, the spectral features of MXB 221120 necessitate specific geometric or physical conditions.

Moreover, we note that the whole burst can be generally described by a single FRED function, and it exhibits an unusually long duration and high kT compared to other MXBs characterized by the same single process, making it an outlier in the burst sample. Given that the FRB-associated MXB is likely triggered in the polar cap region, where strong local magnetic fields lead to tighter confinement of the fireball and consequently result in a higher temperature, MXB 221120 is notable for its emission exhibiting a single process that lasts for an extended duration. This observation suggests that the burst is likely related to a single fracture in the crust from a singular dissipation of internal magnetic energy, rather than multiple fragmented cracks arising from multiple separated triggers.

To summary, we find that both temporal and spectral properties of MXB 221120 observed by GECAM-B indicate that this FRB-associated MXB is very peculiar burst and may be produced by a special mechanism with extreme physical conditions, shedding lights on the origin of MXB as well as FRB.

\begin{acknowledgements}
We acknowledge the support by the National Key R\&D Program of China (2021YFA0718500), HXMT+GECAM
the National Natural Science Foundation of China (Grant Nos. 12273042,
12494572,
12373047, 
12333007
), China's  Space Origins Exploration Program, the Strategic Priority Research Program of the Chinese Academy of Sciences (Grant Nos. XDA30050000,
and XDB0550300
) and the Science Research Project of Hebei Education Department (BJ2026091). 
The GECAM (Huairou-1) mission is supported by the Strategic Priority Research Program on Space Science (Grant No. XDA15360000) of Chinese Academy of Sciences. 
We would like to acknowledge helpful discussions with and Wen-Long Zhang.
\end{acknowledgements}

\clearpage

\bibliographystyle{aasjournal}

\bibliography{main.bib}

\end{document}